\documentclass[twocolumn,showpacs,preprintnumbers,amsmath,amssymb]{revtex4}
\usepackage[dvips]{graphicx}
\usepackage{dcolumn}
\usepackage{bm}    

\begin{document}
\title{Oscillations and synchronization in a system of three reactively
coupled oscillators}
\author{Alexander~P.~Kuznetsov$^1$, Ludmila~V.~Turukina$^{1,2}$, \\
Nikolai~Yu.~Chernyshov$^{3}$
and Yuliya~V.~Sedova$^{1,*}$}
\affiliation{$^{1}$Kotel'nikov's Institute of Radio-Engineering and Electronics of RAS, Saratov Branch,\\
Zelenaya 38, Saratov, 410019, Russian Federation\\
$^2$Department of Physics and Astronomy, Potsdam University, 14476
Potsdam-Golm, Germany\\
$^{3}$ Saratov State University, \\
Astrachanskaya 83, Saratov, 410012, Russian Federation}

\date{\today}

\begin{abstract}
We consider a system of three interacting van der Pol oscillators with
reactive coupling. Phase equations are derived, using proper order of
expansion over the coupling parameter. The dynamics of the system is studied
by means of the bifurcation analysis and with the method of Lyapunov
exponent charts. Essential and physically meaningful features of the
reactive coupling are discussed.

$^{*}$Corresponding author. Tel.:+7 8452278685; fax: +7 8452272401

E-mail address: sedovayv@yandex.ru (Yu.V.Sedova)
\end{abstract}

\pacs{05.45.-a, 05.45.Xt}

\maketitle

\section{Introduction}
Phenomena of synchronization of oscillators are wide spread in nature and
technology. A variety of examples can be found in electronics, laser
physics, biophysics, chemistry, neuroscience, etc. \cite{b1,b2,b3,b4,b5,b6}. Synchronization in
modern experiments for optomechanical, micromechanical, electronic
oscillators is investigated, e.g. see \cite{b7,b8,b9,b10,b11}. The general picture of
oscillatory modes in arrays of elementary oscillators substantially depends
on a number of the elements and of the coupling type. In the simplest case
of two oscillators with \textit{dissipative coupling} one can observe mutual mode locking of the
oscillators with different natural frequencies, two-frequency quasi-periodic
oscillations, the effect of "the oscillation death", and the regime called
the "broadband synchronization" specific for the oscillatory elements with
non-identical control parameters \cite{b1,b2,b3,b12,b13,b14,b15,b16}. With increase of a number of
the oscillatory elements the picture becomes more complicated; many features
of it have been established and understood recently \cite{b17,b18,b19,b20,b21,b22,b23}.

More complex is a case of \textit{reactive coupling} (termed sometimes as the conservative coupling)
\cite{b1,b2,b3,b13,b14,b15}. In radio-engineering and electronics, the coupling of such
kind occurs in the case of presence of a reactive element (inductance) in
the coupling circuit instead of a resistor giving rise to the dissipative
coupling \cite{b2}. A topical example of a system with the reactive coupling
corresponds to ionic traps \cite{b24}. In such traps ions are confined using
variable microwave fields, which restrict a magnitude of radial oscillations
of the ions, and a constant electric field, limiting the axial motions. In a
trap with many electrodes, the ions form a chain being located in the
potential wells, and the nonlinearity provides the anharmonic nature of
oscillations of the ions in the wells. Additionally, the ions are irradiated
by laser beams. The blue laser light of frequency larger than the natural
frequency of the ion oscillations provides instability in the system. The
red laser light of frequency less than the oscillation frequency gives rise
to dissipation. The ions in the chain are coupled due to the Coulomb
repulsion. The simplest model of such a system is a chain of reactively
coupled van der Pol oscillators \cite{b24}. In other concern the models of coupled
van der Pol oscillators are studied in application to biological circadian
rhythms \cite{b25} and for arrays of nanoscale mechanical resonators \cite{b26}.

The reactive coupling is a phenomenon essentially more subtle than the
dissipative coupling. The reason is that when constructing a reduced phase
model one must take into account the second order effects in the coupling
magnitude. In the case of two oscillators in the first order approximation
the coupling effects are generally not manifested. For three or more
oscillators the linear terms are present, but the resulting abridged system
is conservative \cite{b27,b28}.

The case of two oscillators was discussed in Refs. \cite{b13,b14,b15}. There the
appropriate model for dynamics of the phase variable was derived and it was
shown that the synchronization effect appears only with taking into account
the terms of the second order in the coupling parameter. One more feature of
the dynamics with the reactive coupling is the phase bistability; it means
that depending on initial conditions the oscillators may synchronize either
in phase, or in counter-phase. In the present Letter we consider dynamics
and synchronization in a chain of three reactively coupled van der Pol
oscillators. In contrast to \cite{b27,b28}, we will assume non-identical
oscillators for frequency and focus on the discussion of structure of
parameter plane of frequency detuning. The first task is derivation of
correct phase equations accounting all relevant effects. Then, they are
studied using approaches developed in Refs. \cite{b22,b23}. We reveal possible
modes of complete and partial synchronization of the oscillators, the
bifurcation mechanisms of destruction of the synchronization, and describe
arrangement of the parameter domains of quasi-periodic modes with different
number of incommensurable frequencies. One of the main questions we discuss
concerns features and distinctions of the reactive coupling in comparison
with the earlier known results for the dissipative coupling.

\section{The phase model}
Consider a set of equations
\begin{equation}
\label{eq1}
\begin{array}{l}
 \ddot {x} - (\lambda - x^2)\dot {x} + x + \varepsilon (x - y) = 0, \\
 \ddot {y} - (\lambda - y^2)\dot {y} + (1 + \Delta _1 )y + \varepsilon (y -
x) + \varepsilon (y - z) = 0, \\
 \ddot {z} - (\lambda - z^2)\dot {z} + (1 + \Delta _2 )z + \varepsilon (z -
y) = 0. \\
 \end{array}
\end{equation}

\noindent
where $\lambda $ is a parameter controlling intensity of the
self-oscillations, $\Delta _1$ and $\Delta _2$ are the detuning parameters
for the second and the third oscillators, and $\varepsilon$ is the coupling
constant.

If the excitation parameter $\lambda $ is small enough, as well as the
detuning parameters, one can apply the slow-amplitude method for the
analysis of the equations (\ref{eq1}). Using the standard approach \cite{b1,b2}, one can
derive the following equations for the real amplitudes $r_i $ and phases of
the oscillators $\psi _i $ (Landau-Stuart equations):
\begin{equation}
\label{eq2}
\begin{array}{l}
 2\dot {r}_1 = r_1 - r_1^3 - \varepsilon r_2 \sin \theta ,\,\, \\
 2\dot {r}_2 = r_2 - r_2^3 + \varepsilon r_1 \sin \theta - \varepsilon r_3
\sin \varphi ,\,\, \\
 2\dot {r}_3 = r_3 - r_3^3 + \varepsilon r_2 \sin \varphi , \\
 2\dot {\psi }_1 = \varepsilon - \varepsilon\frac{r{ }_2}{r_1 }\cos \theta ,\,\, \\
 2\dot {\psi }_2 = 2\varepsilon + \Delta _1 - \varepsilon \frac{r_1 }{r_2
}\cos \theta - \varepsilon \frac{r_3 }{r_2 }\cos \varphi ,\,\, \\
 2\dot {\psi }_3 = \varepsilon + \Delta _2 - \varepsilon\frac{r_2 }{r_3 }\cos \varphi .
\\
 \end{array}
\end{equation}
Here $\theta = \psi _1 - \psi _2 ,\,\,\varphi = \psi _2 - \psi _3 $ are
relative phases of the oscillators, and the parameters are normalized in
respect to the small parameter $\lambda $ \cite{b14,b15}:
\begin{equation}
\label{eq3}
r \to \sqrt \lambda r,
\quad
t \to t / \lambda ,
\quad
\varepsilon \to \lambda \varepsilon ,
\quad
\Delta \to \lambda \Delta .
\end{equation}
Subtracting pairwise the phase equations (2), we obtain the following
equations for the relative phases:
\begin{equation}
\label{eq4}
\begin{array}{l}
 2\dot {\theta } = - \varepsilon - \Delta _1 + \varepsilon \left( {\frac{r_1
}{r_2 } - \frac{r_2 }{r_1 }} \right)\cos \theta + \varepsilon \frac{r_3
}{r_2 }\cos \varphi , \\
 2\dot {\varphi } = \varepsilon + \Delta _1 - \Delta _2 + \varepsilon \left(
{\frac{r_2 }{r_3 } - \frac{r_3 }{r_2 }} \right)\cos \varphi - \varepsilon
\frac{r_1 }{r_2 }\cos \theta . \\
 \end{array}
\end{equation}

Let us set $r_i = 1 + \tilde {r}_i $, where the tilde designates
perturbations of the stationary orbits $r = 1$. The amplitude equations are
strongly damped \cite{b1,b13,b14,b15}, so the orbits in a short time reach roughly the
stationary amplitudes with some perturbations easily estimated from (\ref{eq2}) as
\begin{equation}
\label{eq5}
2\tilde {r}_1 = - \varepsilon \sin \theta ,\,\,2\tilde {r}_2 = \varepsilon
\sin \theta - \varepsilon \sin \varphi ,\,\,2\tilde {r}_3 = \varepsilon \sin
\varphi .
\end{equation}
In turn, from the phase equations (\ref{eq4}) we obtain
\begin{equation}
\label{eq6}
\begin{array}{l}
 2\dot {\theta } = - \varepsilon - \Delta _1 + 2\varepsilon (\tilde {r}_1 -
\tilde {r}_2 )\cos \theta \\
\phantom {2\dot {\theta } =- \varepsilon - \Delta _1 +2\varepsilon (\tilde {r}_1} + \varepsilon (1 + \tilde {r}_3 - \tilde {r}_2
)\cos \varphi , \\
 2\dot {\varphi } = \varepsilon + \Delta _1 - \Delta _2 + 2\varepsilon
(\tilde {r}_2 - \tilde {r}_3 )\cos \varphi\\
\phantom {2\dot {\theta } =- \varepsilon - \Delta _1 +2\varepsilon (\tilde {r}_1} - \varepsilon (1 + \tilde {r}_1 -
\tilde {r}_2 )\cos \theta . \\
 \end{array}
\end{equation}
Substituting the expressions for the perturbations from (\ref{eq5}) we get
\begin{equation}
\label{eq7}
\begin{array}{l}
 2\dot {\theta } = - \varepsilon - \Delta _1 + \varepsilon \cos \varphi -
\varepsilon ^2\sin 2\theta \\
\phantom {2\dot {\theta } = -}+\varepsilon ^2\left( {\sin \varphi \cos \theta
- \frac{1}{2}\sin \theta \cos \varphi + \frac{1}{2}\sin 2\varphi } \right),
\\
 2\dot {\varphi } = \varepsilon + \Delta _1 - \Delta _2 - \varepsilon \cos
\theta - \varepsilon ^2\sin 2\varphi \\
\phantom {2\dot {\varphi } =-}+ \varepsilon ^2\left( {\sin \theta
\cos \varphi - \frac{1}{2}\sin \varphi \cos \theta + \frac{1}{2}\sin 2\theta
} \right). \\
 \end{array}
\end{equation}
These are the correct phase equations for three reactively coupled
oscillators derived up to the terms of order $\varepsilon ^{2}$. Their
structure is notably more complex than that for the dissipative coupling
\cite{b1,b22}.

In contrast to the case of two oscillators \cite{b13,b14,b15}, the phase equations (\ref{eq7})
do contain terms of the first order in the coupling strength $\varepsilon $,
however, for proper description of the synchronization effects these terms
are not sufficient. Indeed, if one neglect the quadratic terms, the matrix
for perturbations of the stationary state of (\ref{eq7}) is
\begin{equation}
\label{eq8}
\hat {M} = \left( {\begin{array}{l}
 0,\,\,\,\,\,\,\,\,\,\,\,\,\,\, - \varepsilon \sin \varphi \\
 \varepsilon \sin \theta ,\,\,\,\,\,\,0 \\
 \end{array}} \right).
\end{equation}
The trace of this matrix is zero, $S = 0$. It means that a kind of
``neutral'' state occurs on the border between the stable and unstable
solutions (conservative dynamics). Hence, in the system there is no main
resonance at all in this approximation. As follows, for description of
synchronization phenomena one has necessarily take into account the effects
of the second order in the coupling parameter.

Figure 1a illustrates bifurcations of equilibrium points of the system (\ref{eq7}).
In the case of dissipative coupling the border of the domain of complete
synchronization corresponds to a curve of degenerate saddle-node
bifurcations, where simultaneous merging occurs for a pair of the saddles
with the stable and the unstable nodes \cite{b22,b29}. For the reactive coupling
the curves of the saddle-node bifurcations SN for merging of a stable node
and a saddle and for an unstable node and a saddle do not coincide\footnote{
In Fig.1 the curves of unstable node bifurcations are not shown to avoid
cluttering the figure.}. Moreover, here a bifurcation of Andronov-Hopf is
possible (designated by H), where the equilibrium point becomes unstable
with appearance of the stable limit cycle departing from it. Therefore, the
region of complete synchronization in the case of reactive coupling of phase
oscillators appears to be bounded both by the curves of the saddle-node
bifurcations and of the line of the Andronov-Hopf bifurcations. Also we
indicate in Figure 1a the points of codimension two: the cusp point CP and
the Bogdanov--Takens point BT. In Figure 1b, the domains inside of which
system has one or two stable equilibria are shown using different colors.
Thus, there is the simplest multistability.
\begin{figure}[!ht]
\centerline{
\includegraphics[height=9cm, keepaspectratio]{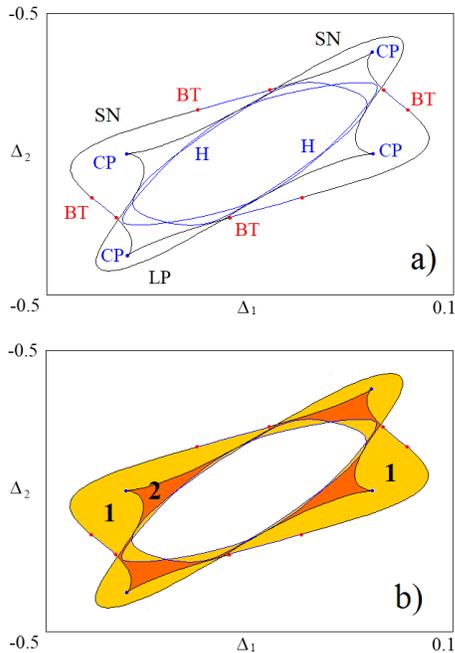}}
\caption{Bifurcation curves and points of the system (\ref{eq7}),
$\varepsilon = 0.2$. Digits in fragment b) indicate the number of coexisting
stable equilibriums.}
\end{figure}

Figure 2 shows the chart of Lyapunov exponents \cite{b22,b23} of the system (\ref{eq7}) on
the parameter plane of frequency detuning of the oscillators ($\Delta
_{1}$,$\Delta _{2})$. To draw the chart we compute two Lyapunov
exponents of the system (\ref{eq7}) $\Lambda _1 $ and $\Lambda _2 $ at each pixel of
the picture and attribute it with a color depending on the signature of the
Lyapunov spectrum to visualize the following regimes:

\noindent
a) $\Lambda _1 < 0,\;\Lambda { }_2 < 0$ -- the complete synchronization of
three oscillators $P $(red),

\noindent
b) $\Lambda _1 = 0,\;\Lambda { }_2 < 0$ -- the two-frequency
quasi-periodicity $T_2 $ (yellow),

\noindent
c) $\Lambda _1 = 0,\;\Lambda { }_2 = 0$ -- the three-frequency
quasi-periodicity $T_3 $ (blue).\\
(Here the types of regimes are responsible to original system (\ref{eq1}). In this
case it is convenient to compare the results obtained at investigations of
the phase model and the original system, see Fig.7.)
\begin{figure}[!ht]
\centerline{
\includegraphics[height=7.5cm, keepaspectratio]{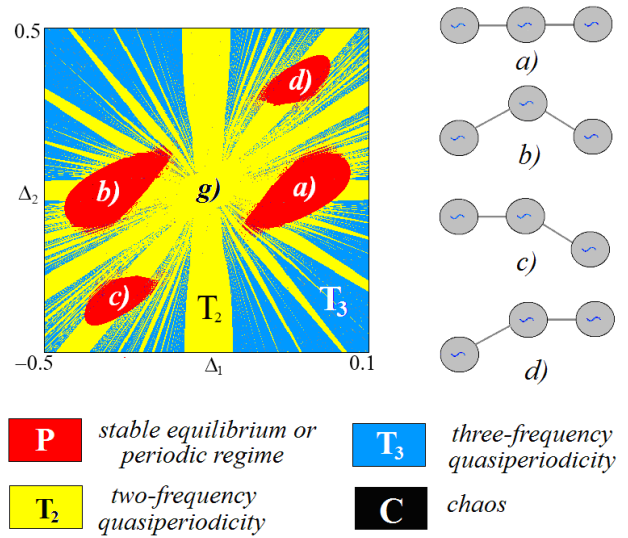}}
\caption{Chart of Lyapunov exponents on the parameter plane and
configurations of basic modes of complete synchronization of the system (\ref{eq7})
at $\varepsilon = 0.2$. The letters designate points corresponding to the
phase portraits of Fig.~3. Used hereinafter color palette for the Lyapunov
charts is shown below.}
\end{figure}

The area of complete synchronization in Fig.~2 contains four
``islands''\footnote{ In fact, in accordance with Figure 1b, the regions of
complete synchronization are overlapped. On Lyapunov charts this is not
visible because of the possibility of multistability in the system.}. In
each island we observe a specific kind of complete synchronization as
illustrated in the phase portraits of Fig.~3 from (a) to (d). At the point
(a) the relative phases are close to zero: $\theta \approx 0,\,\varphi
\approx 0$, and this is the synchronization mode of the \textit{in-phase} type. At the point
(b) the relative phases are $\theta \approx - \pi ,\,\varphi \approx \pi $;
so, the first and the third oscillators are roughly in phase while the
second oscillator is in the counter-phase relatively to them. This is the
\textit{counter-phase} synchronization. In the rest two islands we observe the complete
synchronization of \textit{mixed} type. In this case one of the pairs of the oscillators
(1-2 or 2-3) are in phase, while the rest is in the counter-phase relative
to them. The corresponding configurations of chain are shown in the right
part of Fig.2\footnote{ Analogous modes for coupled Bonhoeffer--van der Pol
oscillators were reported in \cite{b30}.}.
\begin{figure}[!ht]
\centerline{
\includegraphics[height=12cm, keepaspectratio]{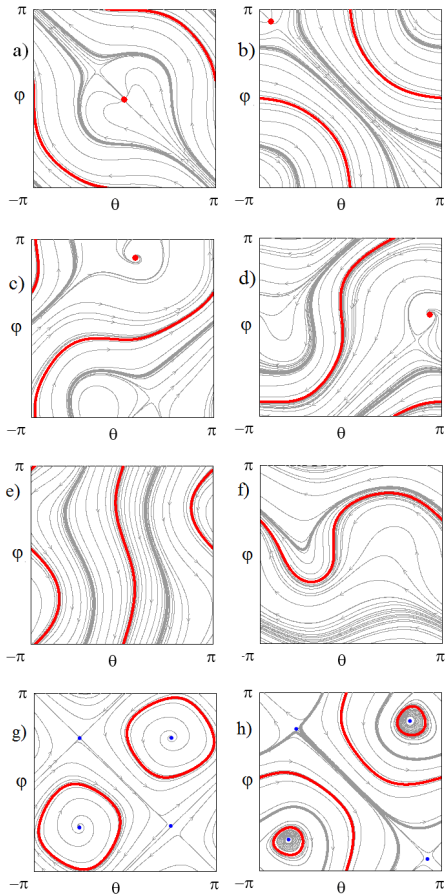}}
\caption{Phase portraits for the system of three reactively
coupled oscillators (\ref{eq7}) at $\varepsilon=0.2$:
a) $\Delta _{1}=0$, $\Delta _{2}=0$; b) $\Delta _{1}=-0.45$, $\Delta
_{2}=0$; c) $\Delta _{1}=-0.38$, $\Delta _{2}=-0.33$; d) $\Delta
_{1}=0$, $\Delta _{2}=0.36$; e) $\Delta _{1}=-0.2$, $\Delta
_{2}=0.4$; f) $\Delta _{1}=0.05$, $\Delta _{2}=0.22$; g) $\Delta
_{1}=-0.2$, $\Delta _{2}=0$; h) $\Delta _{1}=-0.3$, $\Delta
_{2}=0$.}
\end{figure}

Beside the region of complete synchronization, on the Lyapunov chart of
Fig.~2 one can see a set of bands of two-frequency quasi-periodic regimes
immersed in the domain of three-frequency regimes. Within each such band
invariant curves of different types occur in the phase plane. Say, in the
diagram 3(e) the relative phase of the first and the second oscillators
$\theta $ fluctuates around a certain equilibrium value, while the phase
$\phi $ varies across the whole range of values. This is \textit{a partial mode-locking of the first and the second oscillators}. In the diagram
3(f) one observes bounded oscillations of the relative phase of the second
and thirds oscillators $\phi $, so this is \textit{a partial mode-locking of the second and the third oscillators}.

We can classify regions of two-frequency modes with help of the rotation
number $w = p:q$. Here $p$ and $q$ are numbers of intersection of the
corresponding invariant curve with vertical and horizontal sides of the
phase square. Only significant intersections should be used taking into
consideration $2\pi $-periodicity of phase. So for Fig.3e the rotation
number $w = 0:1$ (for both curves) and for Fig.3f $w = 1:0$.

This classification becomes obvious if we compute a ``torus map'' using the
numerical calculation of the factors $p$ and $q$ at each point in the parameter
plane. The corresponding illustration is given in Fig.4. We use the
following rule of coloring. Blue color associates with regime of the
rotation number $w = 1:0$. Decreasing of the rotation number corresponds to
a piecemeal transformation of this color to green mode $w = 1:1$. Then green
color is gradually transformed into the red for the mode with the rotation
number $w = 0:1$. Stable equilibriums are shown in white and the other modes
-- in black. Light gray color corresponds to the regime with contractible
limit cycles when invariance curve has no significant intersections with the
sides of the phase of a square.
\begin{figure}[!ht]
\centerline{
\includegraphics[height=6.5cm, keepaspectratio]{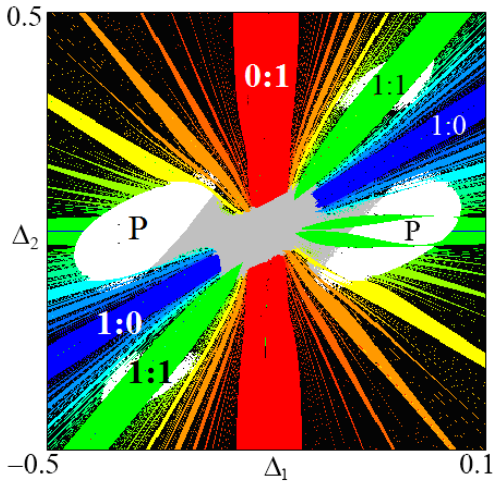}}
\caption{Map of the two-frequency quasi-periodicity area of the
system (\ref{eq7})}
\end{figure}

Origin of the two frequency bands with a common symmetry center in Figure 2
may be explained by the presence of various possible resonances in the
system. To substantiate this point, outline \textit{the partial frequencies} $\omega_i$ of the system
(\ref{eq6}). If we "switch off" in each equation all other oscillators, in the
linear approximation we obtain from (\ref{eq6})
\begin{equation}
\label{eq9}
\omega _1 = \frac{\varepsilon }{2},
\quad
\omega _2 = \varepsilon + \frac{\Delta _1 }{2},
\quad
\omega _3 = \frac{\varepsilon }{2} + \frac{\Delta _2 }{2}.
\end{equation}
A feature of the reactive coupling is that it shifts the partial frequencies
by a value of order $\varepsilon $ (the shift for the central oscillator is
twice as large as for the other ones because it interacts with two rather
than one neighbors).

Now, let us write out the main resonances for the partial frequencies. In
round brackets the respective conditions are indicated in terms of the
frequency detuning parameters:
\begin{equation}
\label{eq10}
\begin{array}{l}
 \omega _1 = \omega _2 \;\left( {\Delta _1 = - \varepsilon } \right),\\
\omega _2 = \omega _3 \;\left( {\Delta _2 = \Delta _1 + \varepsilon }
\right),\\
 \omega _1 = \omega _3 \;\left( {\Delta _2 = 0} \right),\\
  \omega _1 +\omega _3 = 2\omega _2 \;\left( {\Delta _2 = 2\Delta _1 + 2\varepsilon }
\right). \\
 \end{array}
\end{equation}

The resonance conditions (\ref{eq10}) determine the centers for the wide bands of
two-frequency modes in Fig.~2. Additional resonances are possible too, say,
$\omega _1 + \omega _2 = 2\omega _3 $ etc. They correspond to more narrow
bands in Fig.~2. The resonances of higher order give rise to invariant
curves with larger number of intersections with the sides of the phase
square.

Note that the resonances $\omega _1 = \omega _3 $ and $\omega _1 + \omega _3
= 2\omega _2 $ determine lines of symmetry on the parameter plane, see Figs.
1 and 2. This is due to symmetry of these resonance conditions in respect to
permutation of the first and third oscillators. Say, under the condition
$\omega _1 = \omega _3 $, i.e.$\,\Delta _2 = 0$, the equations (\ref{eq7}) are
transformed one to other under the variable change $\theta \leftrightarrow -
\varphi $. As a result, the phase portraits are symmetric about the line
$\varphi = - \theta $, see Fig.~3g,h\textbf{.} Analogously, with the
condition $\Delta _2 = 2(\Delta _1 + \varepsilon )$ the equations are
invariant in respect to the variable change $\theta \leftrightarrow \varphi
+ \pi $.

A case is interesting of equality of all the partial frequencies $\omega _1
= \omega _2 = \omega { }_3$ that corresponds to the symmetry center on the
parameter plane $\Delta _1 = - \varepsilon $,$\Delta _2 = 0$. The phase
portrait at this point is shown in Fig. 3g, and at the point close to it at
- in Fig. 3h. In this case invariant curves of other kind arise, which are
distinct in their topological properties. The curves in Fig.~3g may be
called \textit{contractible}, while those in diagrams ($a)$-($f)$ are called \textit{rotational} \cite{b31} (for the
dissipative coupling the second case is typical \cite{b22}). In the first
case we have a limit cycle going around the unstable equilibrium point.
Then, both the relative phases fluctuate about some mean value. This mode
may be characterized as \textit{a partial mode-locking of all three oscillators}. The frequency spectrum produced by the system (\ref{eq1})
will contain not only the basic frequency, but also a set of components
associated with an additional new time scale, the period of travel of the
representative point around the limit cycle of the phase model.

Contractible limit cycles can occur, as we have already noted, as a result
of Andronov-Hopf bifurcation. In addition, the system can demonstrate
nonlocal bifurcations when the resulting limit cycle arises from the
separatrix loop of the saddle.

Note that several types of multistability are possible in the system with
reactive coupling as seen from Fig.~3. Diagrams ($a)$-($d)$ correspond to
situation when an equilibrium state in the phase space corresponding to the
complete synchronization coexists with an invariant curve corresponding to a
two-frequency quasi-periodic regime\footnote{ Actually, Fig.2 is a
composition of four charts associated with four variants of the initial
conditions and four possible modes of complete synchronization in the
system.}. In diagram ($e)$ two regimes coexist (the in-phase mode and the
counter-phase mode) corresponding to partial mode-locking of the first and
second oscillators. In turn, in diagram ($g)$ there coexist two regimes of
partial synchronization of all three oscillators.

Multistability affects the appearance of the Lyapunov chart. Namely, when
choosing different initial conditions one can observe either periodic or
quasi-periodic regimes, as can be seen from a comparison of Figures 1 and 2.

\section{Dynamics of the original system}
Let us illustrate the effectiveness of the phase model. For this purpose we
represent the results of the bifurcation analysis of the original system (\ref{eq1})
at $\lambda =0.1$, Figure 5. In accordance with the renormalization rules (3) the coupling parameter $\varepsilon = 0.02$ is selected, which allows to compare Figure 5 for the original system and Figure 1 for the phase model. Now instead of a saddle-node bifurcation of
equilibria there is a corresponding bifurcation of limit cycles SNC, instead
of the Andronov-Hopf bifurcation - Neimark-Sacker bifurcation NS, and
instead of Bogdanov-Takens points - points of 1:1 resonance R1. From Figure 1 and Figure 5 we can see that the Figures are similar to each other, which indicates the effectiveness of the phase model for these values of coupling parameter.
\begin{figure}[!ht]
\centerline{
\includegraphics[height=9cm, keepaspectratio]{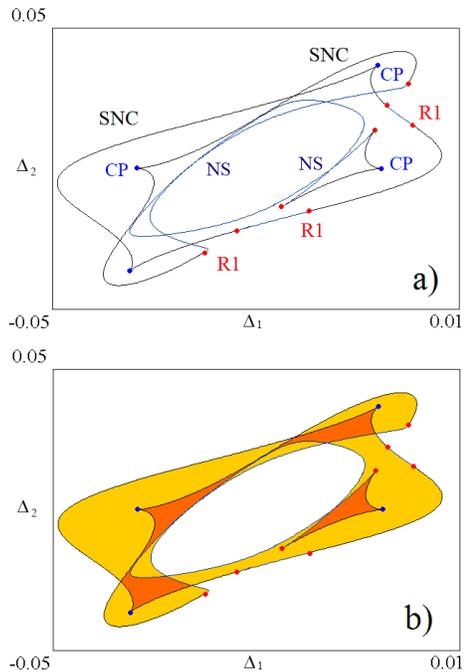}}
\caption{Bifurcation curves and points of the system (1),  $\lambda=0.1$, $\varepsilon=0.02$.}
\end{figure}

It should be noted that the effectiveness of the phase model will increase
with a decrease in the coupling parameter $\varepsilon $. With increasing of
coupling parameter its efficiency falls. Fig.6 illustrates this fact,
demonstrating bifurcation lines of the phase model and the original system
(\ref{eq1}). For the phase model we select coupling parameter $\varepsilon = 0.6$,
so for original system we have $\varepsilon = 0.06$ (taking into account
$\lambda = 0.1$ rescaled on (\ref{eq3})). Some common features - the presence of
four lobes - remain. However, the pictures are different in details. Namely,
external boundaries of lobes now are mainly the Neimark-Saker lines NS.
Lines of saddle-node bifurcations of limit cycles SNC form only small
segments of the boundary of complete synchronization area in the vicinity of
the cusp points CP associated with bistability areas. One more new feature
is the appearance of a double Neimark-Sacker bifurcation points NS-NS.
\begin{figure}[!ht]
\centerline{
\includegraphics[height=9cm, keepaspectratio]{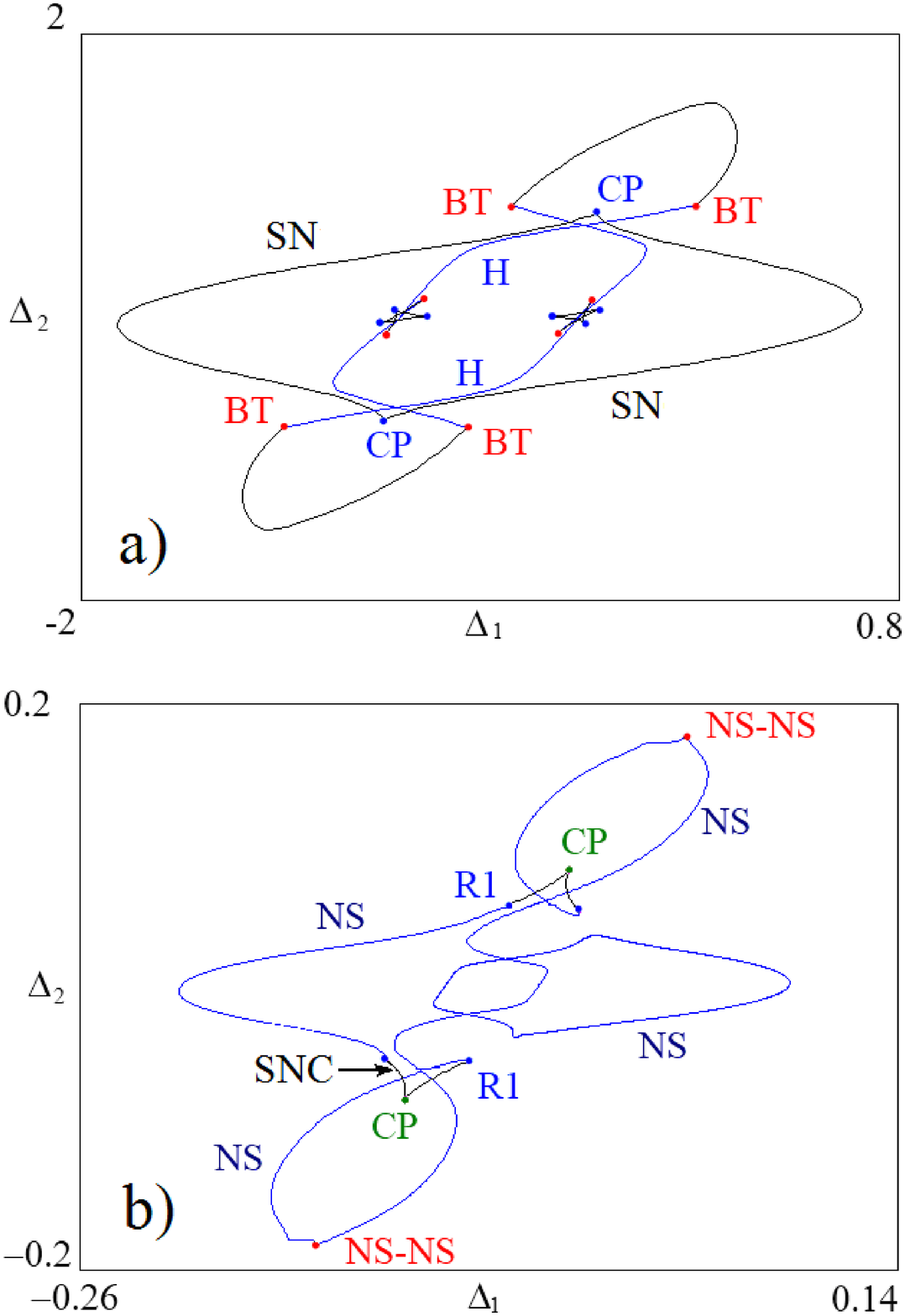}}
\caption{a) Bifurcation lines and dots of phase model (\ref{eq7}),
$\varepsilon = 0.6$; b) the similar illustration for the original system
(\ref{eq1}), $\lambda = 0.1,\varepsilon = 0.06$.}
\end{figure}

Chart of Lyapunov exponents of the original system (\ref{eq1}) is presented in Fig.7
for $\lambda = 0.1,\varepsilon = 0.06$. An enlarged fragment of this chart
in Fig.7b should be compared with Fig.6b.
\begin{figure}[!ht]
\centerline{
\includegraphics[height=11cm, keepaspectratio]{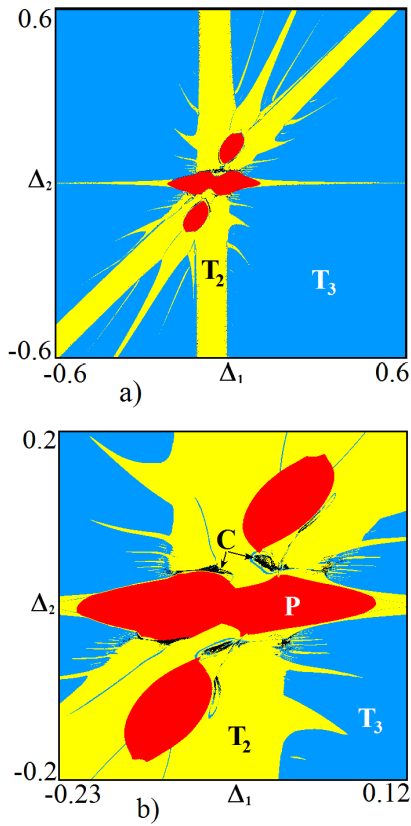}}
\caption{The chart of Lyapunov exponents and its
magnified fragment of the system (\ref{eq1}), $\lambda = 0.1,\varepsilon = 0.06$.}
\end{figure}

With the growth of the $\lambda $ control parameter both the Landau-Stuart
and phase model will work bad to worse. In fig.8a the Lyapunov chart is
shown for the system (\ref{eq1}) for the case $\lambda $=1, $\varepsilon $=0.6. Now
the picture is much more complex. Fig.8b shows a new effect: at the
intersection of numerous bands of two-frequency modes there are located
regions of higher resonances, to which there correspond periodic regimes.
This is a characteristic \textit{resonance Arnold web} \cite{b32}. Letter D denotes the area of the
trajectories' escape to infinity.
\begin{figure}[!ht]
\centerline{
\includegraphics[height=11cm, keepaspectratio]{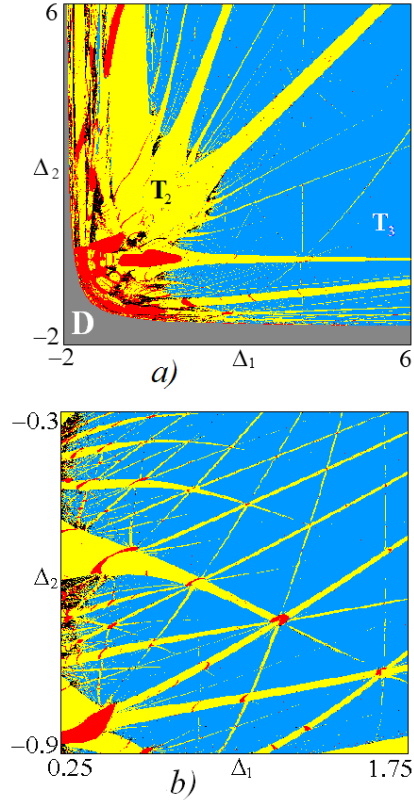}}
\caption{The chart of Lyapunov exponents and its magnified fragment
of the system (\ref{eq1}), $\lambda $=1, $\varepsilon = 0.6$.}
\end{figure}

\section{Conclusion}
The results presented here may be of interest for systems such as ion traps
\cite{b24}, biophysical systems \cite{b25}, etc. At the same time, due to universality
of our models, they are important as well in the general theory of
synchronization. The main result is that in the description of reactive
coupling in the framework of the slow amplitudes approach it is of principal
importance to account the effects of the second order in the coupling
constant. The region of complete synchronization consists of four "islands"
in which the synchronization modes are observed corresponding to in-phase,
counter-phase, and mixed types of oscillations in the chain. The bifurcation
picture of the model formulated in terms of phase variables is significantly
different from the case of dissipative coupling. In particular, Andronov --
Hopf bifurcation is possible responsible for occurrence of partial
synchronization regimes of all three oscillators. For the arrangement of the
parameter plane of the frequency detunings, resonances between the partial
frequencies are important, and for the reactive coupling the resonance
conditions appear to depend on the coupling strength. One more feature is
that the quasi-periodic modes of various types can co-exist with complete
synchronization, giving rise to many kinds of multistability (coexistence of
different attractors).

\section{Acknowledgement}

This work was supported by the grants from the President of the Russian
Federation for support of leading scientific schools NSH-1726.2014.2 and
Russian Foundation for Basic Research project 15-02-02893. Yu.V.S. also
thanks Russian Foundation for Basic Research (grant No.14-02-31064).


\begin{thebibliography}{99}

\bibitem{b1} A. Pikovsky, M. Rosenblum, J. Kurths, Synchronization: A Universal Concept
in Nonlinear Science, Cambridge University Press, 2001.

\bibitem{b2}	P.S. Landa, Nonlinear Oscillations and Waves in Dynamical Systems, Kluwer
Academic Publishers, Dordrecht, 1996.

\bibitem{b3} A.G. Balanov, N.B. Janson, D.E. Postnov, O. Sosnovtseva, Synchronization:
from simple to complex, Springer, 2009.

\bibitem{b4} Y. Kuramoto, Chemical Oscillations, Waves, and Turbulence, New York:
Springer-Verlag, 1984.

\bibitem{b5} L. Glass, M.C. MacKey, From Clocks to Chaos, Princeton University Press,
1988.

\bibitem{b6} A. Winfree, The Geometry of Biological Time, New York: Springer-Verlag, 2001.

\bibitem{b7} G. Heinrich, M. Ludwig, J. Qian, B. Kubala, F. Marquardt, Collective
Dynamics in Optomechanical Arrays, Phys. Rev. Lett. 107 (2011) 043603.

\bibitem{b8} M. Zhang, G. S. Wiederhecker, S. Manipatruni, A. Barnard, P. McEuen, M.
Lipson, Synchronization of Micromechanical Oscillators Using Light, Phys. Rev. Lett. 109 (2012) 233906.

\bibitem{b9} A.A. Temirbayev, Y.D. Nalibayev, Z.Z. Zhanabaev, V.I. Ponomarenko, M.
Rosenblum, Autonomous and forced dynamics of oscillator ensembles with
global nonlinearcoupling: An experimental study, Phys. Rev. E 87 (2013) 062917.

\bibitem{b10} E.A. Martens, S. Thutupalli, A. Fourri\`{e}re, O. Hallatschek, Chimera states in mechanical oscillator networks, Proc. Natl. Acad. Sci. 110(26) (2013) 10563-10567.

\bibitem{b11} M.R. Tinsley, S. Nkomo, K. Showalter, Chimera and phase-cluster states
in populations of coupled chemical oscillators, Nature Phys. 8 (2012) 662-665.

\bibitem{b12} D.G. Aronson, G.B. Ermentrout, N. Kopell, Amplitude response of coupled
oscillators, Physica D 41(3) (1990) 403-449.

\bibitem{b13} R.H. Rand, P.J. Holmes, Bifurcation of periodic motions in two weakly coupled
van der Pol oscillators // Int. J. Non-Linear Mechanics 15(4-5) (1980) 387-399.

\bibitem{b14} M.V. Ivanchenko, G.V. Osipov, V.D. Shalfeev, J. Kurths, Synchronization of two
non-scalar-coupled limit-cycle oscillators, Physica D 189(1-2) (2004) 8-30.

\bibitem{b15} A.P. Kuznetsov, N.V. Stankevich, L.V. Turukina, Coupled van der Pol--Duffing
oscillators: Phase dynamics and structure of synchronization tongues, Physica D 238(14) (2009) 1203-1215.

\bibitem{b16} A.P. Kuznetsov, Ju. P. Roman, Properties of synchronization in the systems
of non-identical coupled van der Pol and van der Pol--Duffing oscillators.
Broadband synchronization, Physica D 238(16) (2009) 1499-1506.

\bibitem{b17} P. Ashwin, J. Guaschi, J.M. Phelps, Rotation sets and phase-locking in an
electronic three oscillator system, Physica D 66(3-4) (1993) 392-411.

\bibitem{b18} P. Ashwin, Boundary of two frequency behaviour in a system of three weakly
coupled electronic oscillators, Chaos, Solitons and Fractals 9(8) (1998) 1279-1287.

\bibitem{b19} P.S. Linsay, A.W. Cumming, Three-frequency quasiperiodicity, phase locking,
and the onset of chaos, Physica D 40(2) (1989) 196-217.

\bibitem{b20} P. Ashwin, O. Burylko, Y. Maistrenko, Bifurcation to heteroclinic cycles and
sensitivity in three and four coupled phase oscillators, Physica D 237(4) (2008) 454-466.

\bibitem{b21} Yu. Maistrenko, O. Popovych, O. Burylko, P.A. Tass, Mechanism of
desynchronization in the finite-dimensional Kuramoto model, Phys. Rev.
Lett. 93 (2004) 084102.

\bibitem{b22} Yu.P. Emelianova, A.P. Kuznetsov, I.R. Sataev, L.V. Turukina,
Synchronization and multi-frequency oscillations in the low-dimensional
chain of the self-oscillators, Physica D 244(1) (2013) 36-49.

\bibitem{b23} Y.P. Emelianova, A.P. Kuznetsov, L.V. Turukina, I.R. Sataev, N.Yu.
Chernyshov, A structure of the oscillation frequencies parameter space for
the system of dissipatively coupled oscillators, Communications in Nonlinear
Science and Numerical Simulations 19(4) (2014) 1203-1212.

\bibitem{b24} T.E. Lee, M.C. Cross, Pattern formation with trapped ions, Phys. Rev.
Lett. 106 (2011) 143001.

\bibitem{b25} K. Rompala, R. Rand, H. Howland, Dynamics of three coupled van der Pol
oscillators with application to circadian rhythms, Communications in
Nonlinear Science and Numerical Simulation 12(5) (2007) 794-803.

\bibitem{b26} M.C. Cross, A. Zumdieck, R. Lifshitz, J.L. Rogers, Synchronization by
nonlinear frequency pulling, Phys. Rev. Lett. 93 (2004) 224101.

\bibitem{b27} A. Pikovsky, P. Rosenau, Phase compactons, Physica D 218(1) (2006) 56-69.

\bibitem{b28} P. Rosenau, A. Pikovsky, Phase compactons in chains of dispersively coupled
oscillators, Phys. Rev. Lett. 94 (2005) 174102.

\bibitem{b29} V. Anishchenko, S. Astakhov, T. Vadivasova, Phase dynamics of two coupled
oscillators under external periodic force, Europhysics Letters 86(3) (2009) 30003.

\bibitem{b30} A.K. Kryukov, G.V. Osipov, A.V. Polovinkin, J. Kurths, Synchronous regimes in
ensembles of coupled Bonhoeffer-van der Pol oscillators, Phys. Rev. E 79 (2009) 046209.

\bibitem{b31} C. Baesens, J. Guckenheimer, S. Kim, R.S. MacKay, Three coupled oscillators:
mode locking, global bifurcations and toroidal chaos, Physica D 49(3) (1991) 387-475.

\bibitem{b32} H. Broer, C. Sim\'{o}, R. Vitolo, The Hopf-saddle-node bifurcation for fixed
points of 3D-diffeomorphisms: the Arnol'd resonance web, Bull. Belg. Math. Soc. Simon Stevin 15(5) (2008) 769-787.

\end{thebibliography}
\end{document}